\documentclass[journal,twocolumn]{IEEEtran}
\usepackage{amsfonts}
\usepackage{times}
\usepackage{graphicx}
\usepackage{epstopdf}
\usepackage{latexsym}
\usepackage{dsfont}
\usepackage{amssymb}
\usepackage{amsmath}
\usepackage{cite}
\usepackage{verbatim}
\usepackage{subfigure}

\def\bb0{{\mathbb{0}}}

\def\bb{{\mathbf{b}}}

\def\bff{{\mathbf{f}}}

\def\bh{{\mathbf{h}}}

\def\bp{{\mathbf{p}}}

\def\bs{{\mathbf{s}}}

\def\b0{{\mathbf{0}}}

\def\bX{{\mathbf{X}}}

\def\bbC{{\mathbb{C}}}

\def\bbE{{\mathbb{E}}}

\def\bbP{{\mathbb{P}}}

\def\bbR{{\mathbb{R}}}

\def\bbZ{{\mathbb{Z}}}

\def\cF{\mathcal{F}}

\def\cO{\mathcal{O}}

\def\sf0{{\mathsf{0}}}

\usepackage{graphicx}
\usepackage{epstopdf}
\usepackage{enumerate}
\usepackage{algorithm}
\usepackage{amsmath}
\usepackage{mathrsfs}
\usepackage{float}
\usepackage{color}
\usepackage{makeidx}
\usepackage{bm}
\usepackage{cleveref}
\usepackage{url}
\usepackage{steinmetz}
\usepackage{varwidth}
\usepackage{soul}
\usepackage{bbm}
\usepackage{empheq}
\usepackage{mathtools}
\usepackage{cite}
\usepackage{balance}

\newcommand{\sref}[1]{{Section}~\ref{#1}}

\usepackage{amsfonts}
\usepackage{booktabs}
\usepackage{siunitx}

\begin{document}
\title{Computer Vision Aided Beam Tracking in \\ A Real-World Millimeter Wave Deployment}
\author{Shuaifeng Jiang and Ahmed Alkhateeb\\ Arizona State University - Emails: \{s.jiang, alkhateeb\}@asu.edu \thanks{The authors are with the School of Electrical, Computer and Energy Engineering, Arizona State University, Tempe, AZ, 85281 USA (Email: s.jiang, alkhateeb@asu.edu). This work is supported by the National Science Foundation under Grant No. 2048021.}}

\maketitle
\begin{abstract}
Millimeter-wave (mmWave) and terahertz (THz) communications require beamforming to acquire adequate receive signal-to-noise ratio (SNR). To find the optimal beam, current beam management solutions perform beam training over a large number of beams in pre-defined codebooks. The beam training overhead increases the access latency and can become infeasible for high-mobility applications. To reduce or even eliminate this beam training overhead, we propose to utilize the visual data, captured for example by cameras at the base stations, to guide the beam tracking/refining process. We propose a machine learning (ML) framework, based on an encoder-decoder architecture, that can predict the future beams using the previously obtained visual sensing information. Our proposed approach is evaluated on a large-scale real-world dataset, where it achieves an accuracy of $64.47\%$ (and a normalized receive power of $97.66\%$ ) in predicting the future beam. This is achieved while requiring less than $1\%$ of the beam training overhead of a corresponding baseline solution that uses a sequence of previous beams to predict the future one. This high performance and low overhead obtained on the real-world dataset demonstrate the potential of the proposed vision-aided beam tracking approach in real-world applications.
\end{abstract}
\section{Introduction}
The millimeter-wave(mmWave) and terahertz (THz) have been considered as key enabler for the high data rate communication in future wireless networks \cite{rappaport2019wireless}. The high carrier frequencies provide an order of magnitude more bandwidth compared with existing wireless communication systems. The move to higher frequencies, however, brings new challenges such as the higher path-loss. To overcome that and ensure sufficient receive power, mmWave/sub-THz communication systems need to deploy large antenna arrays at the transmitters/receivers and use narrow beams. Nevertheless, obtaining the optimal narrow beams often requires large beam training overhead, which occupies wireless resources and decreases spectral efficiency. This becomes more significant for high-mobility applications such as autonomous vehicles and vehicle-to-everything (V2X) communications \cite{V2X}, which are considered key applications for future wireless communication systems \cite{advanced_app}. All that motivates the need to develop novel approaches that can find the optimal beams with low or negligible beam training overhead.
\par
An important observation is that the use of narrow beams at mmWave/sub-THz networks and the reliance on line-of-sight (LoS) links give a special importance to the knowledge of the physical location of the transmitters/receivers and the geometry of the environment around the communication systems. This motivates the use of position/environment sensing devices (such as position sensors, cameras, etc.) at the communication terminals to guide the different link establishment/resource allocation tasks. Prior works have studied improving mmWave/THz beam selection, blockage prediction, and hand-over based on sensing information of different modalities \cite{alrabeiah2020deep, handoff, charan2021visionposition, beamblockage, identification, demirhan2021radarbeam, chou2021fast}. In \cite{alrabeiah2020deep} the author proposes that the sub-6 GHz channel contains adequate information of mmWave channel, therefore it can be used to predict mmWave beam and blockage status. The vison/camera sensing modality has been increasingly studied \cite{handoff, beamblockage, identification}. \cite{handoff} leverages the camera signals to proactively predict dynamic link blockages and handoff for mmWave systems. \cite{  beamblockage} employs cameras at the mmWave base stations and leverages the visual data to help overcome the beam training overhead. In \cite{charan2021visionposition, chou2021fast, demirhan2021radarbeam}, other data modalities such as the radar senory information and the UE position are utilized to guide the beam selection process. However, the above work focused mainly on the current beam prediction task instead of the beam tracking which aims to predict future beams. Beam tracking is particularly important for the sensing-aided mmWave communicatons systems as the latency of capturing and processing the sensory data will unlikely enable current beam prediction, but could be useful for predicting future beams.
\par
In this paper, we propose to utilize visual sensing information to enable fast and low-overhead mmWave/THz beam tracking. The main contribution can be summarized as follows.
\begin{itemize}
\item We propose a universal problem formulation for the auxiliary data-aided beam tracking, which could be used for different auxiliary data modalities, such as leveraging a sequence of beams, or a sequence of RGB images.
\item We propose a machine learning (ML) framework for sensing information aided mmWave/THz beam tracking exploiting an encoder-decoder architecture.
\item We evaluate the proposed vision-aided beam tracking on the real-world DeepSense 6G dataset \cite{DeepSense}, which comprises co-existing visual and wireless beam data.
\end{itemize}
Simulation results demonstrate the capability of the proposed vision-aided beam tracking approach in achieving high accuracy and receive power. This highlights the potential gains of incorporating visual sensors (cameras) in real-world mmWave/THz communication systems.
\section{System and Problem Formulation}\label{System and Problem Formulation}
In Section \ref{System and Problem Formulation}, we first introduce the considered system model for mmWave communications. Then, we formulate beam tracking into an optimization problem. After that, we clearly define the vision-aided beam tracking machine learning task. Lastly, we also present a baseline beam tracking ML task using the previous optimal beam sequence.
\subsection{System Model}
\begin{figure}[t]
\centering
\includegraphics[width=0.9\linewidth]{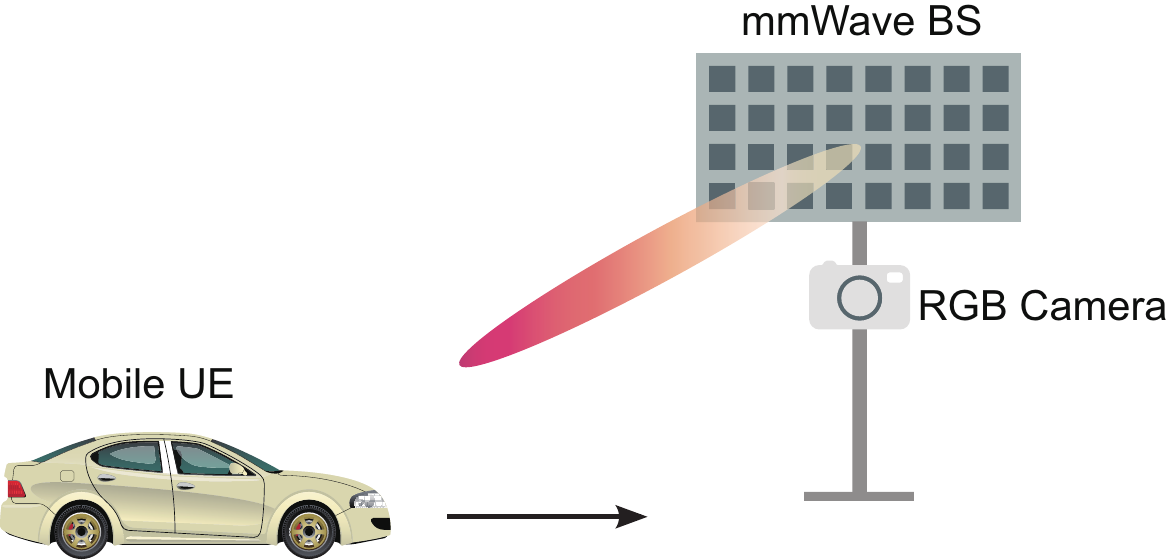}
\caption{This figure illustrates the considered system model: the BS senses the environment and the moving UE with an RGB camera. The obtained sensing information is then utilized for the BS beam management.}
\label{fig:system_model}
\end{figure}

Fig. \ref{fig:system_model} shows the considered system model for mmWave communications, where the base station (BS) is serving a mobile user equipment (UE). The BS is equipped with an antenna array of $N$ elements and an RGB camera (visual data sensor). Using the antenna array, the BS performs beamforming to achieve adequate receive power. We assume that the BS has a pre-defined beamforming codebook ${\boldsymbol{\cF}} = \{\bff_1,\hdots, \bff_{|{\boldsymbol{\cF}}|}\}$ containing $|{\boldsymbol{\cF}}|$ beams $\bff_m \in \bbC^{N \times 1}$ \cite{Zhang2021}. For the sake of simplicity, the UE is assumed to have a single antenna. At time step $t$, the BS transmits a compelex symbol $s[t]\in \bbC$. We assume the downlink signal $s[t]$ satiesfies the power constraint $\bbE\left[s^Hs\right]=P$ with $P$ denoting the transmit power. Then, the corresponding downlink receive signal $y[t]$ can be written as
\begin{align}\label{eq:signal model}
y[t] = \bh^H[t] \bff[t] s[t] + n[t],
\end{align}
where $\bh[t] \in \bbC^{N \times 1}$ denotes the channel between the BS and the UE at time step $t$. $\bff[t] \in {\boldsymbol{\cF}}$ is the transmit beamforming vector used at the BS at time step $t$. $n[t]$ is the receive noise which satisfies $\bbE[n[t] n^H[t]] = \sigma_n^2$, and $\sigma_n^2$ denotes the receive noise power.
\subsection{Problem Formulation}
This paper focuses on the beam tracking problem at the base station, which is defined as follows: Given the available sensing inforamtion up to time $t-1$, the BS attempts to determine the optimal beams of $\xi \in \bbZ^+$ future time steps, that is, optimal beams of $t, \hdots, (t+\xi-1)$. First, we define the {\bf optimal beam} at time step $t$ as the one which gives the {\bf highest beamforming gain}. The optimal beam at time step $t$ is then represented by
\begin{align}\label{eq:bbest beam}
\bff^\star[t] &=\underset{\bff[t] \in {\boldsymbol{\cF}}}{\arg\max}\ \big|\bh^H[t] \bff[t]\big|^2.
\end{align}
With this pre-defined codebook constraint, the optimal beam $\bff^\star[t]$ can be uniquely represented by its beam index in the codebook. The optimal beam index at time step $t$ satisfies
\begin{align}\label{eq:best idx}
p^\star[t] = \underset{p[t] \in [1,|{\boldsymbol{\cF}}|]}{\arg\max}\ \big|\bh^H[t] \bff_{p[t]} \big|^2.
\end{align}
Note that, under the codebook constraint, obtaining the optimal beam is equivalent to obtaining the optimal beam index. With the definition of the optimal beam index in \eqref{eq:best idx}, we formulate the beam tracking problem as follows:
\begin{equation} \label{eq:optip}
\begin{aligned}
\max_{\hat{p}[t]} \quad & \bbP\left\{\hat{p}[t] = p^\star[t] \mid \cO_{t-\xi} \right\}\\
\textrm{s.t.} \quad & \hat{p}[t] \in [1, |{\boldsymbol{\cF}}|],\\
\end{aligned}
\end{equation}
where $\hat{p}[t]$ is the predicted optimal beam index for time step $t$. $\cO_{t-\xi}$ is any auxiliary obtained {\bf before} time step \mbox{$(t-\xi+1)$} that contains the partial information of the optimal beam at time step $t$.
\par
\begin{figure*}[t]
	\centering
	\includegraphics[width=0.9\linewidth]{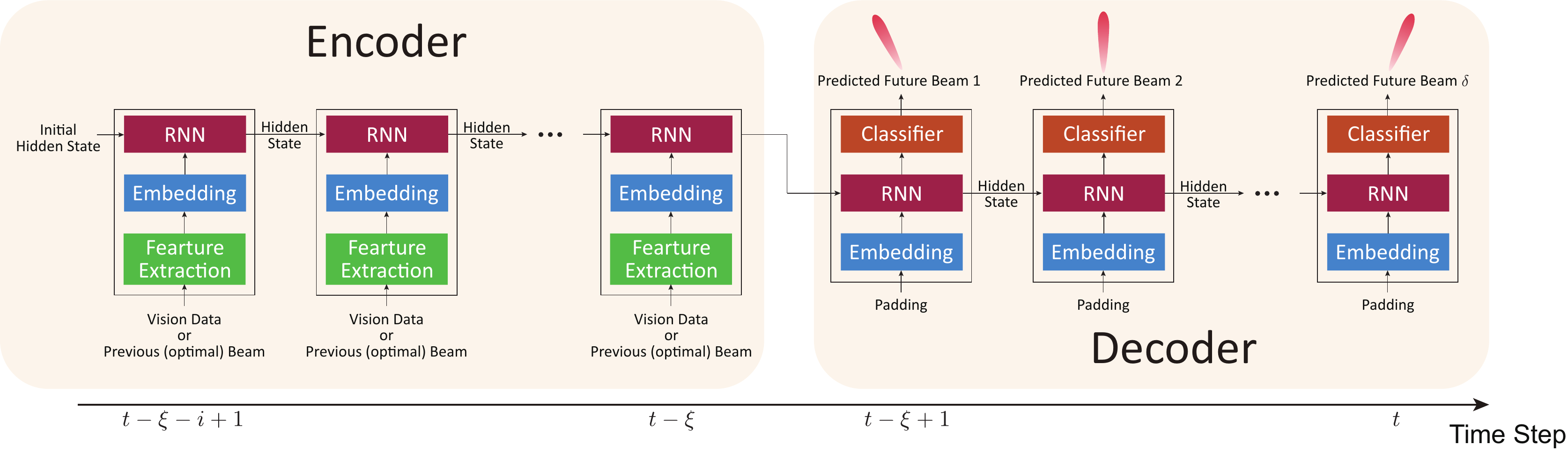}
	\caption{This figure shows the block diagram of the proposed ML framework for sensing aided beam tracking. The ML framework adopts an encoder-decoder architecture, and incorporates the feature extraction block, the embedding block, the RNN block, and the classifier block.}
	\label{fig:mlframework}
	\vspace{-0.1cm}
\end{figure*}

\subsection{vision-aided Beam Tracking}
In this paper, We propose exploiting the visual sensing inforamtion to achieve accurate beam tracking. The BS uses its RGB camera to capture the visual sensing information of the mobile UE and the surrounding environment. Let $\bX[t] \in \bbR^{H \times W \times 3}$ denote the visual sensing information (RGB image) captured at time step $t$. $H$ and $W$ are the height and width of the captured image, and the last dimension represnts the $3$ RGB channels. To utilize this visual sensing information for beam tracking, our objective then becomes obtaining a function which can predict the optimal future beams starting from time step $t$ based on the visual sensing information obtained up to time step $t-1$. Let $X_{t, i} = \left\{\bX[t-i+1],\hdots, \bX[t]\right\}$ denote a sequence of visual sensing information with $i$ representing the observation time window. Then, from \eqref{eq:optip}, the vision-aided beam tracking optimization problem can be wirtten as
\begin{equation} \label{eq:optip_v}
\begin{aligned}
\max_{\hat{p}[t]} \quad & \bbP\left\{\hat{p}[t] = p^\star[t] \mid X_{t-\xi, i} \right\}\\
\textrm{s.t.} \quad & \hat{p}[t] \in [1, |{\boldsymbol{\cF}}|]\\
\end{aligned}
\end{equation}
Since the precise joint probability distribution of $p[t]^\star$ and $X_{t-\xi, i}$ is difficult to model, we propose to leverage the powerful learning capabilitites of ML models to solve \eqref{eq:optip_v} in an data-driven approach. Let $f(;\theta)$ denote an ML model with $\theta$ representing its trainable parameters. To solve the vision-aided beam tracking in \eqref{eq:optip_v}, the ML model aims at predicting the optimal beam index $p^\star[t]$ using the side information $X_{t-\xi, i}$. Therefore the optimal ML model for vision-aided beam tracking can be mathmatically represented by
\begin{align}\label{eq:optm2}
f_v^\star(;\theta_v^\star) = \underset{f_v(;\theta_v)}{\arg\max} \ \bbP\left\{f_v\left(X_{t-\xi,i};\theta\right)=p^\star[t]\right\},
\end{align}
where $\theta_v^\star$ is the associated optimal parameters of $f^\star$.
\subsection{Baseline Beam Tracking}
The sequence of beams resulting from exhaustive search beam training in the previous time steps may also carry information of the mobile UE's future optimal beam. Therefore, it can be exploited as the side information $\cO_{t-\xi}$ in \eqref{eq:optip} to predict the future beam (in the beam tracking problem). In this paper, we employ this approach as a baseline for the beam tracking task. If this approach is implemented using machine learning, then we define this ML task as:
\begin{align}\label{eq:optm3}
f_{b}^\star(;\theta_{b}^\star) = \underset{f_{b}(;\theta_{b})}{\arg\max} \ \bbP\left\{f_{b}\left(F^\star_{t-\xi,i};\theta_{b}\right)=p^\star[t]\right\},
\end{align}
where $F^\star_{t,i} = \left\{ \bff^\star[t-i+1],\hdots,\bff^\star[t] \right\}$ denotes the optimal beam sequence from time step $(t-i+1)$ to time step $t$. $f_{b}^\star$ and $\theta_{b}^\star$ are the optimal ML model and trainable parameters associated to this baseline beam tracking task which uses the previous optimal beam sequence as input.
\par

In Section \ref{Proposed Soulution}, we will explain in detail the proposed ML models for the vision-aided beam tracking and the baseline beam tracking tasks.
\section{Vision-Aided Beam Tracking}\label{Proposed Soulution}
In this section, we first elaborate on the motivation of exploiting sensing to overcome the difficulty of mmWave communications. Then we propose an ML framework that solves the vision beam tracking and the baseline beamforming tasks formulated in \sref{System and Problem Formulation}.
\subsection{Enabling Highly-Mobile mmWave Systems with Cameras} \label{Enabling mmWave Communications: Sensing with Vision}
mmWave/sub-THz communication is considered as a key component of wireless systems in 5G and beyond. These systems can take advantage of their larger transmission bandwidth to achieve high data rates. The use of high-frequency carriers, however, brings new challenges that originated from the propagation characteristics at these bands. For example, these systems need to use large antenna arrays and narrow beams to achieve sufficient receive signal power. Adjusting these beams, however, involves high beam training overhead \cite{3gpp}. This training overhead becomes even more challenging for high-mobility scenarios such as autonomous vehicles and V2X communications. These are considered as the key applications of advanced communication systems \cite{advanced_app}.
\par
Further, the high propagation loss at mmWave/sub-THz makes these systems highly rely on line-of-sight (LoS) links and dominant non-line-of-sight (NLOS) propagation paths corresponding to the main reflectors. Therefore, the optimal beam for mmWave/THz systems is highly dependent on the direction/position of the transmission targets and the geometry and layout of the surrounding environment. This motivates leveraging sensors, such as cameras \cite{handoff, identification, beamblockage}, to obtain awareness about the user location and the surrounding environment. Using cameras has the following advantages.
\begin{itemize}
\item Camera provides rich and fine-grain information on the potential transmission target and the surroundings.
\item Camera does not require any wireless resource in its operation (compared to radars), and can also work without feedback signals (compared with wireless communication based sensing).
\item Cameras are mature and well-commercialized sensors. They have the advantage in terms of price, and are also relatively easy for implementations and deployments.
\item It is possible to utilize the advanced models and algorithms developed for computer vision applications.
\end{itemize}
\par
Next, we will present the proposed ML framework for the sensing aided beam tracking task, and explain in detail the DL model proposed for the vision-aided beam tracking.
\subsection{Developed Deep Learning Model}
Fig. \ref{fig:mlframework} shows the block diagram of the proposed ML framework for the sensing-aided beam tracking task. The ML framework adopts an encoder-decoder architecture featuring four components: the feature extraction block, the embedding block, the RNN block, and the classifier block. The encoder processes the previously obtained information, and passes the information to the decoder. The decoder predicts the future beams based on the information it receives from the encoder.
\par
\textbf{Fearture Extraction Block}: The first component of the proposed ML framework is the feature extraction block which directly processes the raw input data. The raw input data often contain extra information which does not contribute to the beam tracking tasks. This irrelevant information can be detrimental since the ML model can overfit on them while neglecting the useful information. Moreover, dropping this unnecessary information also results in a smaller feature space, thus, results in a more stable training process and lower computational overhead. Therefore, the feature extraction block is designed to filter out the unwanted information and extract the features that are informative for the downstream task.
\par
\begin{figure*}[t]
	\centering
	\includegraphics[width=1\linewidth]{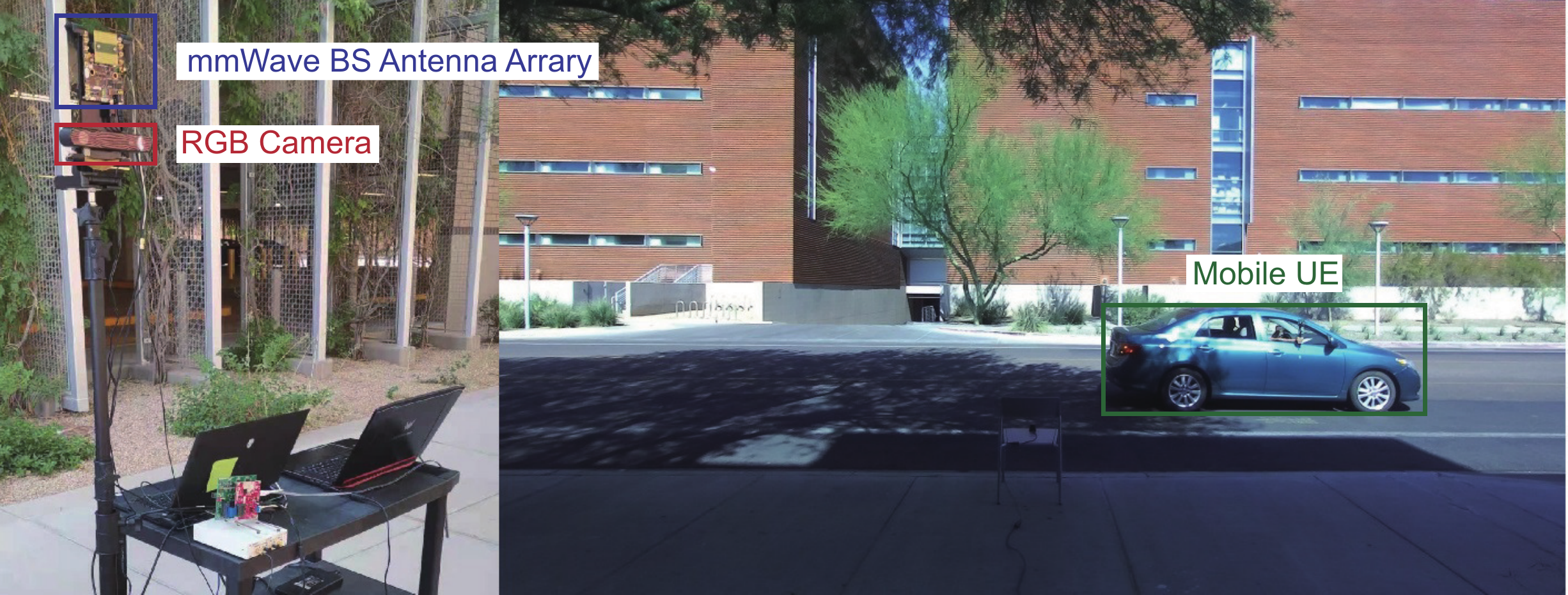}
	\caption{System setup of the DeepSense 6G Secnerio 8. The mmWave BS antenna arrays opterateing at 60GHz receives the signal transmitted by the moving UE. A Camera is placed under the BS to obtain sensing information.}
	\label{fig:setup}
\end{figure*}

To process the raw RGB data for the vision-aided beam tracking task, we detect the potential transmission target (UE) in the RGB image. We take advantage of the advanced computer vision ML models and employ the YOLOv4 \cite{yolov4} object detector. YOLOv4 is a state-of-the-art convolutional neural network (CNN) based ML model designed to detect thousands of classes of objects from real-world RGB images. The YOLOv4 object detector can achieve real-time prediction and high accuracy which is suitable for the vision-aided beam tracking task. Given an input RGB image, the YOLOv4 object detector predicts a class index $c\in\bbZ$, a confidence score $s_c\in [0,1]$, and a bounding box vector $\bb\in\bbR^{4 \times 1}$. Note that the bounding box vector $\bb = [x_c, y_c, w, h]^T$ consists of the $x$-center, $y$-center, width, and height of the detected object in the RGB image. For more detailed information related to the YOLOv4 model, we refer the readers to \cite{yolov4}. As discussed in \sref{Enabling mmWave Communications: Sensing with Vision}, the optimal beam is highly dependent on the direction/position of the transmission target. Therefore, we exploit the bounding boxes as the extracted feature for the successive blocks in the framework.
\par
The feature extraction block is skipped in the baseline approach since the inputs are the optimal beam indices.
\par
\textbf{Embedding Block}: The embedding layer transforms the input feature into a different vector space. The embedded vector ideally captures some of the semantics of the input vector such that semantically similar input vectors form clusters in the embedding space. For the vision feature (the bounding boxes) embedding, we employ a fully connected layer which linearly transforms the bounding box vector $\bb\in \bbR^{4\times 1}$ into $\tilde{\bb}\in \bbR^{E_v\times 1}$. To embed the previous beam indices, we apply the same approach as the natural language processing (NLP) ML models embed the word token. For the $|{\boldsymbol{\cF}}|$ beam indices in the codebook ${\boldsymbol{\cF}}$, we employ a look-up table of $|{\boldsymbol{\cF}}|$ trainable embedding vectors $\left\{ \tilde{\bp}_1,\hdots,\tilde{\bp}_{|{\boldsymbol{\cF}}|} \right\}$. In the simulations, we set $E_v$ and $E_b$ to $64$.
\par
\textbf{RNN Block:} The third component of the proposed ML model is the recurrent neural network (RNN) block. The recurrent neural networks (RNNs) have been extensively studied for processing sequential signal and data such as the NLP and speech recognition tasks. Empirical results have shown that RNNs can effectively capture and process sequential features.
Due to the sequential nature of the beam tracking task, we employ the RNN architecture to process the sequential input $\cO_{t-i}$ and predict the future beams. We adopt single-layer gated recurrent units (GRUs) with $\xi$ units to process the sequence of visual information or the previous optimal beam index. The hidden state size of the GRU is set to 64.
\par
\textbf{Classifier Block:} A fully connected layer is used as the classifier block. This block predicts the future optimal beam index from the high-level features obtained by the RNN block. The softmax activation function is applied to this fully connected layer to output a confidence score vector $\hat{\bs} = [s_1, \hdots, \bs_{|{\boldsymbol{\cF}}|}]^T$ of the beam indices in ${\boldsymbol{\cF}}$. The beam index with the highest score is predicted as the optimal future beam\vspace{-0.1cm}
\begin{align}
\hat{p} =\underset{p \in [1,|{\boldsymbol{\cF}}|]}{\arg\max}\ \bs_p.
\end{align}\vspace{-0.1cm}
\par
\textbf{Learning Phase:} The encoder of the proposed ML framework processes the previously obtained information. The input sequence to the encoder is $X_{t-\xi, i}$ for the vision-aided beam tracking task or $F_{t-\xi, i}$ for the baseline beam tracking task. Based on the information received from the encoder, the decoder predicts the future optimal beams. A padding sequence of $\xi$ zero vectors $Z_\xi$ is input to the decoder as a placeholder.
\par
Since the ML framework is designed to solve a classification problem, we employ the cross-entropy loss. The loss function can be written as \vspace{-0.2cm}
\begin{align}\label{eq:loss1}
J = \sum_{j=t-\xi+1}^{t}\sum_{m=1}^{|{\boldsymbol{\cF}}|}p^\star_{m}[j]\log_2\left(\hat{s}_{m}[j] \right),
\end{align}
where $p^\star_{m}[j]$ is the $m$-th element of the one-hot coded vector of $\bp^\star$ at time step $j$. $\hat{s}_{m}[j]$ is the $m$-th element of the output vector $\hat{\bs}[j]$ at time step $j$.

\section{Experimental Setup}
Our proposed vision-aided beam tracking approach and the ML framework are designed to manage real-world mmWave beam tracking. Therefore, we need a high-quality real-world dataset consisting of co-existing RGB images and beam data to evaluate our proposed approach. In this paper, we adopt the DeepSense 6G dataset \cite{DeepSense} in our simulation and performance evaluation. The DeepSense 6G is a \textbf{multi-modal} dataset comprising \textbf{real-world} measurements. The DeepSence 6G dataset incorporates co-existing including wireless beam data, visual sensing data, among other modalities.
\subsection{DeepSense 6G Secnerio 8}
We adopt Scenario 8 of the DeepSense 6G dataset for our simulation. The system setup of Scenario 8 is shown by \mbox{Fig. \ref{fig:setup}}. In Scenario 8, a fixed BS receives signals from a moving UE. The BS is equipped with a square uniform planar array (UPA) of 64 ($8\times8$) elements and an RGB camera installed under the UPA. The BS adopted a beamforming codebook consisting of $64$ pre-defined beams. The codebook is horizontal-only and $8$-time oversampled. The UE is a moving vehicle equipped with the same UPA and transmits mmWave signal on the 60 GHz band using an omni-directional beam. During the data collection process, the UE passes by the BS multiple times. At each time step, the BS measures the receive power of all beams in the codebook by beam sweeping, and captures the UE with the RGB camera.
\begin{figure}[t]
	\centering
	\includegraphics[width=1\linewidth]{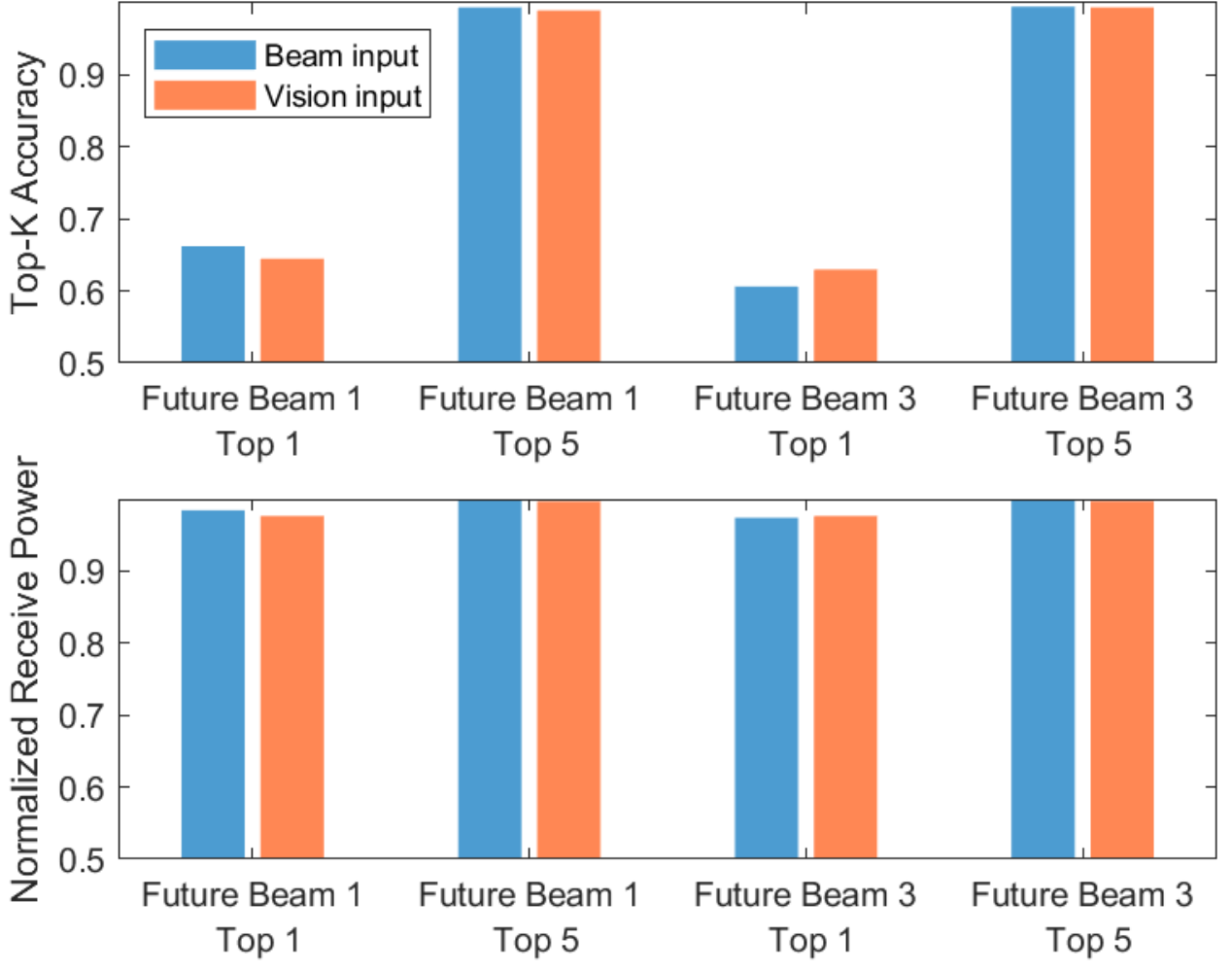}
	\caption{This figure compares the accuracy and normalized receive power of the top-$k$ predictions of future beam 1 and 3. For both the vision-aided and baseline beam tracking, the accuracy increases as $k$ increases. However, the normalized receive power saturates to the optimum even when $k=1$.}
	\label{fig:accpwr}\vspace{-0.3cm}
\end{figure}

\subsection{Devolop Dataset Generation}
We evaluated our proposed vision-aided beam tracking approach on the official development dataset split of DeepSense 6G Scenario 8. The training dataset and the validation dataset consist of $70\%$ and $20\%$ of the raw dataset.
The DeepSence 6G Scenario 8 dataset consists of multiple data sequences. In a data sequence, the vehicle completes its path passing by the BS for one time. Each data sequence consists of co-existing RGB images and beam receive powers of multiple time steps. We follow the official development dataset split of DeepSense 6G Scenario 8. The training dataset and the validation dataset consist of $70\%$ and $20\%$ of the raw dataset data sequences. Note that the data in two datasets come from different vehicle passes to assure there is no data leakage. For each data sequence, we break it into data samples using a sliding window size of $13$. One data sample consists of $13$ time steps and can be written as $\left\{ (\bX[1], p^\star[1]),\hdots,(\bX[13], p^\star[13])\right\}$.
\par
In the training process, we use an observation window size $i=8$, and we train the models to predict the future beam $1\sim5$ ($\xi\in[1,5]$). Therefore, the input to the encoder is $\left\{ \bX[1],\hdots,\bX[8]\right\}$ for the vision-aided beam trakcing model, and $\left\{ p^\star[1],\hdots,p^\star[8]\right\}$ for the baseline beam trakcing model. In both two beam tracking approaches, the decoder is expected to output $\left\{ \hat{\bs}[9],\hdots,\hat{\bs}[13]\right\}$.
\section{Evaluation Results}
In this section, we evaluate the proposed vision-aided beam tracking approach and compares its performance with the baseline beam tracking's performance. The metrics adopted in the evaluation are the following.
\begin{itemize}
\item Top-$k$ accuracy: the percentage of the time steps of all validation samples where the beam corresponding to the top-$k$ confidence scores include the optimal beam.
\item Normalized receive power: the ratio between the highest receive power achieved by the top-$k$ predicted beams and the receive power of the optimal beam. This metric is averaged over all time steps and all validation samples.
\end{itemize}
\par

\subsection{Do the ML Models Learn to Predict Future Beams?}\label{Vision-aided vs. Baseline: Beam Tracking Accuracy}
\begin{figure}
	\centering
	\includegraphics[width=1\linewidth]{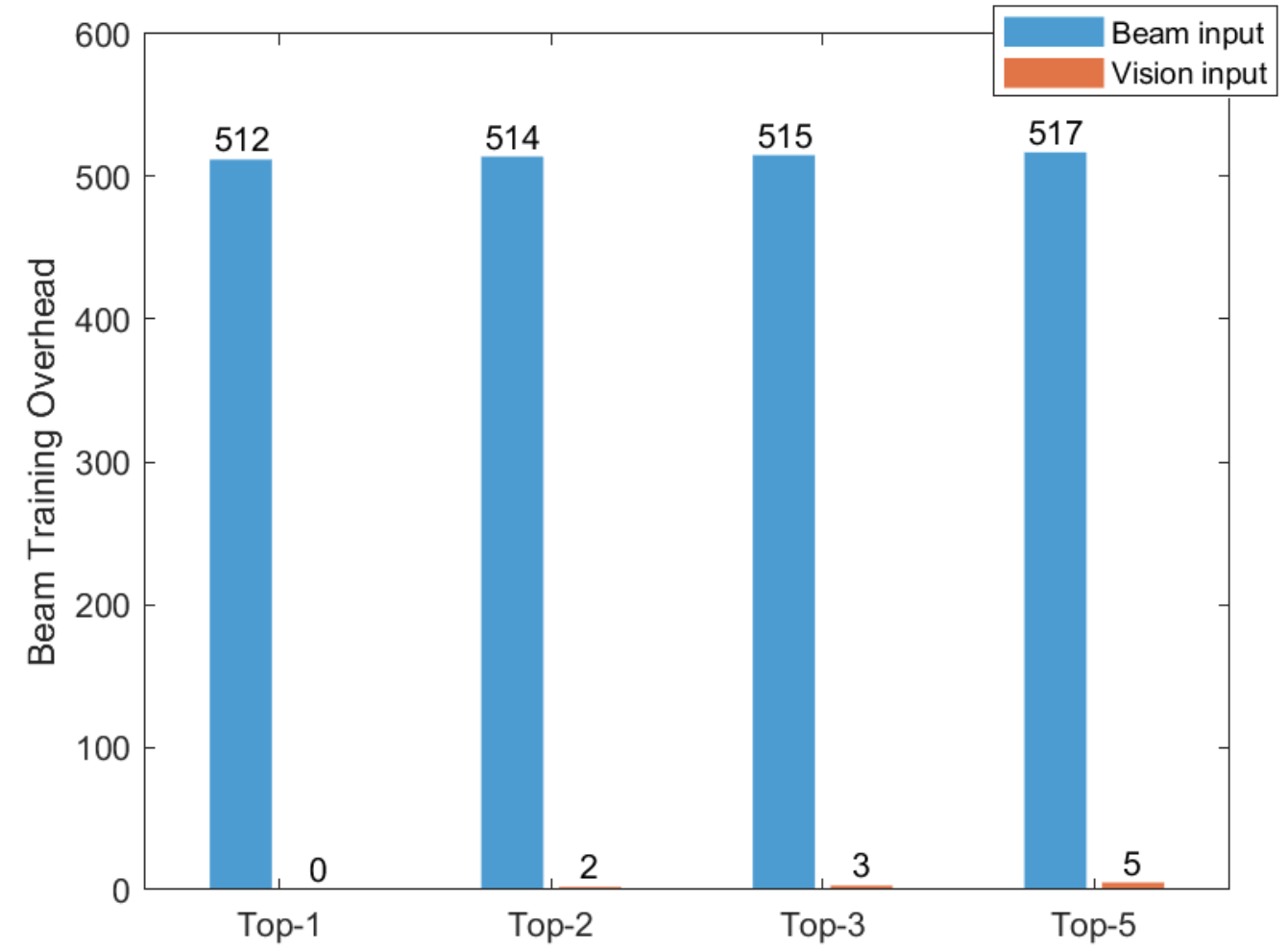}
	\caption{This fugure compares the training overhead required in nine times steps of the vision-aided and the baseline approaches. The vision-aided approach only consumes $1\%$ beam training overhead compared to the baseline.}
	\label{fig:training_overhead_ratio}\vspace{-0.3cm}
\end{figure}

In Fig. \ref{fig:accpwr}, we present the top-$k$ accuracy and normalized receive power of the future beam 1 and future beam 3. The top-$k$ accuracy improves significantly as $k$ increases for both the vision-aided and baseline beam tracking approaches. It can also be observed that accuracy decreases when predicting beams in the further future. The \mbox{top-$5$} accuracy of the vision-aided beam tracking for \mbox{future beam 1} is $99.37\%$. This accuracy implies that the proposed vision beam tracking approach can find the optimal \mbox{future beam 1} with $99.37\%$ probability by testing $5$ beams in the beam sweeping process. However, in terms of receive power, testing more beams (increasing $k$) may not be worth the price of the training overhead. For the vision-aided beam training, \textbf{$\bf 97.66\%$ receive power can already be obtained for future beam 1 even without any beam training ($k=1$)}, leaving little room for testing 5 beams for improvement. Despite the relatively low top-$1$ accuracy obtained by both beam tracking approaches, the near-optimal receive power highlights that the \textbf{ML models effectively learn to predict future beams}, and most of the mistakes occur at the sub-optimal beam with near-optimal receive power. Overall, \textbf{the vision-aided beam tracking approach can achieve comparable performance to the baseline approach in terms of the two metrics.} Note that the baseline beam tracking approach is a strong baseline since it inputs the optimal beams of the $8$ previous time steps. This highlights the capability of the proposed vision-aided approach in accurate beam tracking. It is worth mentioning that the adopted DeepSense 6G scenario 8 mainly consists of LoS data. However, the baseline beam tracking approach could be more sensitive to NLoS scenarios since the previous optimal beam may not contain enough information on the blockages, reflectors, and scatterers. The baseline beam tracking is also expected to further degrade when the surrounding environment becomes more dynamic. On the contrary, The visual data obtained by the camera captures rich information on the surrounding object and the dynamics. Moreover, the baseline beam tracking approach requires information on the optimal previous beams, which may not always be applicable in practice. The baseline model can instead rely on the beams it previously predicted. This may cause the baseline approach to deviate more from the optimal beam as the beam tracking goes on without calibration. The vision-aided approach, however, keeps capturing the latest information on the environment. Therefore, it is not likely to suffer from this deviation.
\subsection{Vision-aided vs. Baseline: Beam Training Overhead}
In \sref{Vision-aided vs. Baseline: Beam Tracking Accuracy}, we analyzed the accuracy and received power performance of the two beam tracking approaches. However, in our simulation, the baseline approach requires knowing the previous eight optimal beams to predict the optimal beam of the future time step and match the performance of the vision-aided approach. Therefore, Fig. \ref{fig:training_overhead_ratio} studies the beam training overhead required by the vision-aided and the baseline approaches in these nine time steps when top-$k$ beams are predicted. It is assumed that both approaches will conduct beam sweeping over the predicted top-$k$ beams at the future time step when $k \geq 2$. \textbf{For the 5 cases shown in Fig. \ref{fig:training_overhead_ratio}, the beam training overhead required by the vision-aided approach is less than $1\%$ of the baseline approach.} Furthermore, when top-$1$ beam is predicted the vision-aided approach completely eliminates the beam training overhead.

\subsection{What is the effect of the Beamforming Codebook Size}
\begin{figure}[t]
\centering
\includegraphics[width=1\linewidth]{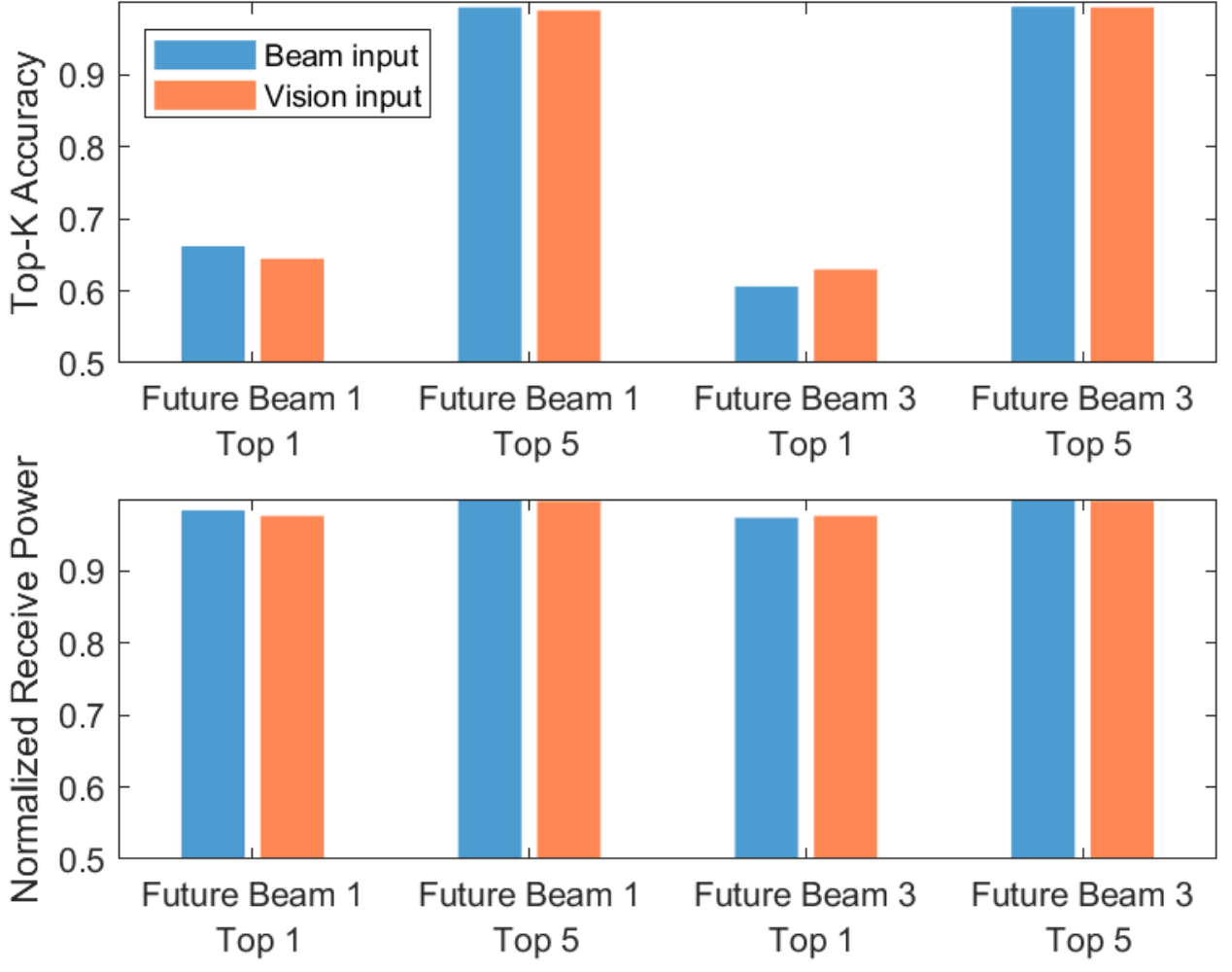}
\caption{This figure shows the accuracy and receive power performance of the vision-aided and baseline beam tracking approaches with different codebook sizes of 16, 32, and 64.}
\label{fig:diffbeam}\vspace{-0.3cm}
\end{figure}

In this section, we study the effect of the codebook size on the performance of the vision-aided and baseline beam tracking approaches. Fig. \ref{fig:diffbeam} shows the top-1 accuracy and normalized receive power of the future beam 1. It can be seen that, as codebook size increases, the top-$k$ accuracies of both beam tracking approaches decrease as can be expected. However, the normalized receive power increases as a larger beam codebook size is adopted. Using the vision-aided beam tracking approach with 16 pre-defined beams, the normalized receive power of the top-$1$ prediction is $92.17\%$ for future beam 1. Exploiting the codebook with $64$ beams, a normalized receive power of $97.66\%$ can be achieved, which is a $6\%$ relative improvement. This demonstrates that \textbf{reasonable receive power improvement can be achieved by using oversampling codebook} at a price of slightly more computational complexity.
\section{Conclusion}
This paper proposes a machine learning based vision-aided beam tracking framework. Exploiting this framework, we also develop an efficient baseline beam tracking approach that utilizes the previous optimal beams. The proposed approaches are evaluated using a large-scale real-world dataset comprising co-existing visual and wireless mmWave data. Evaluation results demonstrate that the proposed vision-aided beam tracking approach can learn to accurately predict future beams and achieve comparable performance to the baseline solution. It achieves a top-1 accuracy of $64.47\%$ and a top-5 accuracy of $98.95\%$ in predicting the future beam. The robustness of the proposed vision-aided beam tracking is illustrated by the $97.66\%$ normalized receive power of the top-$1$ prediction. Moreover, the proposed vision-aided beam tracking only requires $1\%$ of the beam tracking overhead of the baseline approach. These results highlight the potential of leveraging visual sensors in improving mmWave communications.

\balance

\end{document}